\documentclass[submission,copyright,creativecommons]{eptcs}
\usepackage{listings}
\usepackage[english]{babel}
\usepackage{hyperref}
\usepackage{graphicx}
\usepackage{latexsym}
\usepackage{verbatim}
\usepackage{xcolor}
\usepackage{color}
\usepackage{multirow}
\usepackage{url}
\usepackage{booktabs}
\usepackage{algpseudocode}
\usepackage{hyperref}

\newtheorem{theorem}{Theorem}
\newtheorem{property}{Property}
\newtheorem{proposition}{Proposition}

\newenvironment{proof}[1][Proof]{\begin{trivlist}
\item[\hskip \labelsep {\bfseries #1}]}{\end{trivlist}}

\algdef{SE}[EVENT]{Event}{EndEvent}[1]{\textbf{upon event}\ #1\ \algorithmicdo}{\algorithmicend\ \textbf{event}}
\algtext*{EndEvent}
\algdef{SE}[RECEIVE]{Receive}{EndReceive}[1]{\textbf{on receive}\ #1\ \algorithmicdo}{\algorithmicend\ \textbf{receive}}
\algtext*{EndReceive}
\algdef{SE}[SEND]{Send}{EndSend}[1]{\textbf{send}\ #1\ textbf{to}\algorithmicdo}{\algorithmicend\ \textbf{send}}
\algtext*{EndSend}
\newcounter{pseudocode}

\newcommand{\qed}{$\Box$}

\newcommand{\tuple}[1]{\langle{#1}\rangle}

\newcommand{\remove}[1]{}
\title{Model Checking Paxos in Spin}
\author{Giorgio Delzanno\institute{DIBRIS, Universit\`a di Genova}
\and Michele Tatarek\institute{DIBRIS, Universit\`a di Genova} 
\and Riccardo Traverso\institute{FBK Trento}
}
\def\titlerunning{Model Checking Paxos in Spin}
\def\authorrunning{G. Delzanno, M. Tatarek, and R. Traverso}
\begin{document}
%
\maketitle
\begin{abstract}
We present a formal model of a distributed consensus algorithm in the executable specification language Promela
extended with a new type of guards, called counting guards, needed to implement transitions that depend 
on majority voting. Our formalization exploits abstractions that follow from reduction theorems applied
to the specific case-study.
We apply the model checker Spin to automatically validate finite instances of the model and to 
extract preconditions on the size of quorums used in the election phases of the protocol.
\end{abstract}
\section{Introduction}
Distributed algorithms are a challenging class of case-studies for automated verification methods 
(see e.g. \cite{AWN12,BPZ06,SWJ08,AWN12,FHM07,JK08,SRS09,KVW12,DT13,graphite14}).
The main difficulties come from typical assumptions taken in these algorithms such as asynchronous communication media, 
topology-dependent protocol rules, and messages with complex structure.
In the subclass of fault tolerant distributed protocols there are additional aspects to consider that often 
make their validation task harder.  Indeed, fault tolerant protocols  are often based on dynamic leader elections in which quorums may change from one round to another.
When modeling these protocols, one has to deal with a very fast growth of the search space for increasing number of 
processes.

Following preliminary evaluations described in \cite{graphite14}, in this paper we apply Promela/Spin to specify and validate a fault tolerant distributed consensus protocols for asynchronous systems caled Paxos \cite{Paxos,PaxosMadeSimple} used in the implementation of distributed services in the Google File System \cite{Chubby}.
Promela is a specification language for multithreaded systems with both shared memory and communication capabilities.
Promela provides a non-ambiguous executable semantics that can be tested by using a simulator
and the Spin model checker.  Spin applies enumerative techniques for validating finite-automata representation of  Promela models.  

In the paper we give a formal Promela specification of the Paxos protocol that is modular with respect to roles, rounds, and communication media
inspired to the presentation given by Marzullo, Mei and Meling in \cite{PaxosMadeSimple} via three separate roles (proposer, acceptor, and learner). 
The resulting specification is closer to a possible implementation than sequential models with non-deterministic assignments  used in 
other approaches such as \cite{tsuchiya2008using}.

Via a formal analysis extracted from the correctness requirements,
which are specified using auxiliary variables and assertions, we give reduction theorems that can be used to 
restrict the number of processes instances for some of the protocol roles, namely proposers and learners.
Finally, we design code to code Promela transformations that preserve the interleavings from one model to the other 
while optimizing the state exploration process required in the model checking phase.
For this purpose, we introduce a special type of atomic transitions, called quorum transition, that can directly 
be applied to model elections via majority voting. 
The transitions are defined on top of a special guard that counts the number of messages in a channel 
embedded into Promela code using the deterministic step constructor $d\_step$.

\section{Paxos: An Informal Specification}\label{paxos}
The consensus problem requires agreement among a number of agents for some data value. 
Initially, each agent proposes a value to all other ones. 
Agents can then exchange their information. 
A correct agent must ensure that when a node takes the final choice, 
the chosen value is the same for all correct agents.
In the asynchronous setting, it is assumed  that messages can be delayed arbitrarily. 
Furthermore, a subset of processes may crash anytime and restore their state (keeping their local information) after 
an arbitrary delay. Fisher, Lynch and Patterson have shown that, under the above mentioned assumptions, 
solving consensus is impossible \cite{FLP85}.

In \cite{Paxos} Lamport proposed a (possibly non terminating) algorithm, called Paxos, addressing this problem.
Lamport later provided a simpler description of the protocol in \cite{PaxosMadeSimple} in terms of three separate agent roles: 
proposers that can propose values for consensus, acceptors that accept
a value among those proposed, and learners that learn the chosen value.
Marzullo, Mei and Meling give a pseudo-code presentation of the algorithm in \cite{MMH13}.

The Paxos protocol for agreement on a single value works as follows.
In a first step the proposer selects a fresh round identifier and broadcasts it to a (subset of) acceptors.
It then collects votes for that round number from acceptors.
Acceptor's replies, called {\it promises}, contain the round number, and a pair consisting of the last round and value that they
accepted in previous rounds (with the same or a different proposer).
When the proposer checks that majority is reached, it selects a value to submit again to the acceptors.
For the selection of the value, the proposer inspects every promise received in the current round
and selects the value with the highest round. It then submits the current round and the chosen value to the acceptors.
Acceptors wait for proposals of round identifiers but accept only those that are higher than the last one they have seen so far.
If the received round is fresh, acceptors answer with a promise not to accept proposals with smaller round numbers.
Since messages might arrive out-of-order, even different proposals with increasing rounds of the same proposer might arrive 
with arbitrary order (this justifies the need of the promise message).
Acceptors also wait for accept messages:
in that case local information about the current round are updated and, if the round is fresh, 
the accepted pair $(round,value)$ is forwarded to the learner.
A learner simply collects votes on pairs $(round,value)$ sent by acceptors and waits to detect majority for one of them.

The protocol is guaranteed to reach consensus with $n$ acceptors up to $f=\lfloor (n-1)/2\rfloor$ simultaneous failures, 
but only if learners have enough time to take a decision (i.e. to detect a majority).
If proposers indefinitely inject new proposals the protocol may diverge.
Under the above mentioned condition on $f$ and in absence of byzantine faults, correctness can be formulated as follows.
\begin{property}\label{safetyprop}\label{safety}
When a value is chosen by a learner, then no other value has been chosen/accepted in previous 
rounds of the protocol.
\end{property}
This means that, whenever a value is chosen by the learner, any successive voting always select the same value (even with larger round identifiers), i.e.,  the algorithm stabilizes w.r.t. the value components of tuples sent to the learner.
To ensure this property, in the first part of the protocol the proposer does not immediately send its proposal for the value but only its round number.  The proposer selects a value only after having acquired knowledge on the values accepted in previous
rounds.  Among them the choice is done by taking the value of the highest round. Its own proposal comes into play only if
all values received by acceptors are undefined (equal to $-1$).
Other safety requirements are that chosen values are among those proposed and that 
chosen values (by the whole system) are the same as those learned by learners.

\section{A Formal Model in Promela}
In this section we present a formal specification of Paxos given in Promela, a specification language for 
multithreaded and distributed programs.
Promela thread templates are defined via the \verb+proctype+ construct.
The body of the template (a sequence of commands) defines the behavior of its instances.
The language has a C-like syntax for shared and local data structures and instructions.
Guarded commands are used to model non-deterministic choices in the body of process templates.
For instance, the command $\verb+if+\ ::g_1 \rightarrow c_1;\ \ldots\ ::g_n \rightarrow  c_n;\ \verb+fi+$ specifies 
a non-deterministic conditional: only one command $c_i$ among those for which the guard $g_i$ is 
enabled is selected for execution.
The guarded command $\verb+do+\ ::g_1 \rightarrow  c_1;\ \ldots\ ::g_n \rightarrow  c_n;\ \verb+od+$ is the iterative version of the conditional.
Data structures include basic data types (byte, bool, int) as well as arrays and structures.
Furthermore, channels can be used for the specification of inter-process communications.
For instance, $\verb+chan+\ c[MAX]$, where $MAX$ is a constant, defines a channel with at most $MAX$ places.
A message $\tuple{m_1,\ldots,m_n}$ is sent by using the command $c!m_1,\ldots,m_n$, 
where $c$ is a channel and $m_1,\ldots,m_n$ are expressions whose current values are sent on the channel.
Reception is defined via the capability $c?x_1,\ldots,x_n$, where $c$ is a channel and $x_1,\ldots,x_n$ are
variables used to store the data from the incoming message.
Channels can be viewed as global arrays.
The selector \verb+?+ provides FIFO access, whereas \verb+??+  provides random access.
To restrict reception to a given pattern, it is possible to put either a constant value in a reception 
or an expression like $eval(x)$ that evaluates to the current value of $x$.
We describe next our first Promela model for Paxos.

A round is defined via a proposer process running in parallel with the other processes 
(other proposers, acceptors and learners). 
The proposer \verb+proctype+ takes in input two parameters: a unique round identifier (a number) and the proposed value. 
Round identifiers must be unique for the algorithm to work. 
Roles can be viewed as threads definitions within the same process situated in a location.
Asynchronous communication is modeled via global channels with random receive actions.
We model the majority test via conditions on counters that keep track of the number of received messages
(we do not model message duplication).
Since the protocol assumes that each round is associated to a unique proposer we can use a local counter for votes 
received by a given proposer.
Rounds are also used as a sort of time-stamps by acceptors.
Indeed they are required in order to fix an order on incoming \verb+prepare+ and \verb+accept+ messages, 
i.e.,
enforce some order in a chaotic flow of messages.
We manipulate messages inside atomic actions. 
A special learner process is used to observe the results of the handshaking between
proposers and acceptors, and  to choose pairs \verb+round,value+.
The algorithm guarantees that once a value is chosen, such a choice remains stable
when other (old/new) proposals are processed by the agents.
The learner keeps a set of counters indexed on rounds to check for majority on a value.
We discuss next the Promela specification in full detail.
First of all, we use the constants:
\begin{lstlisting}
#define ACCEPTORS 3
#define PROPOSERS 5
#define MAJ (ACCEPTORS/2+1)// majority
#define MAX (ACCEPTORS*PROPOSERS)
\end{lstlisting}
The former defines an upper bound on the size of channels.
The latter defines the size of quorums.
By changing MAJ we can infer preconditions on the number of faulty processes.

Using a thread-like style, we consider four shared data structures that represent 
communication channels. They correspond to different phases of the protocol.
The message signature is defined as follows.
\begin{lstlisting}
chan prepare =  [MAX] of { byte,  byte };
chan accept  =  [MAX] of { byte,  byte,  short };
chan promise =  [MAX] of { mex };
chan learn   =  [MAX] of { short, short, short };
\end{lstlisting}
The proposer of round $r$ sends messages \verb+prepare(i,r)+ and \verb+accept(i,r,v)+ 
to acceptor $i$. Acceptor $i$ sends the message \verb+promise(r,hr,hval)+ to the proposer of round $r$ 
and \verb+learn(i,r,v)+ to the learner. All channels are treated as multiset of messages using the 
random receive operations \verb+??+.

The protocol makes use of broadcast communication (from proposers to acceptors).
We implement a derived broadcast primitive using Promela macro definition via 
the inline declaration. 
Specifically, we add the \verb+baccept+ and \verb+bprepare+ primitives defined as follows,
where $i$ is an integer index local to a process proctype.
\begin{lstlisting}
inline baccept(round,v){
    for(i:1.. ACCEPTORS){ 
       accept!!i,round,v; 
    }
    i=0;
}

inline bprepare(round){
    for(i:1.. ACCEPTORS){ 
        prepare!!i,round; 
    }
    i=0;
}
\end{lstlisting}
In the above listed definitions for each process identifiers
we insert the message specified in the parameter resp. in the
\verb+accept+ and \verb+prepare+ channels. 
The typedef $mex$ will be used later to inspect the content of a channel
using a for-loop (Promela allows this kind of operations on channels 
in which messages have a predefined type).

We now move to the specifications of the protocol roles.
%

%
\paragraph{Proposer Role}
A single round of a proposer is defined via the \verb+proctype proposer+.
defined as follows:
\begin{lstlisting}
proctype proposer(int round; short myval) {
    short hr = -1, hval = -1, tmp;
    short h, r, v;
    byte count;
    bprepare(round);
    do
     :: rec_p(round,count,h,v,hr,hval);
     :: send_a(round,count,hval,myval,tmp);
     od }
\end{lstlisting}
where the atomic transition \verb+rec_p+ is defined as
\begin{lstlisting}
rec_p(round,count,h,v,hr,hval)=
  d_step {
         promise??eval(round),h,v ->
         if :: count < MAJ -> count++; 
            :: else fi;
         if :: h > hr -> 
                 hr = h; 
                 hval = v
            :: else fi; 
         h = 0; v = 0;
    }
\end{lstlisting}
and the atomic transition \verb+send_a+ is defined as
\begin{lstlisting}
send_a(round,count,hval,myval,tmp)=
  d_step{
         count >= MAJ ->
         if :: hval<0 -> tmp = myval
            :: else   -> tmp = hval  fi;
        }
        baccept(round,tmp);
        break
\end{lstlisting}       
The round is identified by a unique value passed as a parameter to the  \verb+proctype+.
The other parameter, \verb+myval+, is the proposed text/value of the proposer.
To generate several proposals, it is necessary to create several instances of the proposer \verb+proctype+
running in parallel. Together with the use of unordered channels, our definition models arbitrary delays 
among the considered set of proposals (old proposals can overtake new ones).
We assume here that proposers are instantiated with distinct round values.
Inside the proctype we use different local variables.
In particular, \verb+count+ is used to count votes (i.e. promise messages) 
and to check if  a majority has been reached in the current round.
\verb+hr+ and \verb+hval+ are local variables used to store resp. the max round identifier seen so far
in \verb+promise+ messages and the associated value.
Variables \verb+h,r,v+ are used to store values of received messages.
The subprotocol starts with a \verb+bprepare+ invocation.
It is used to broadcast \verb+prepare+ messages to all other processes (or to a quorum).
The broadcast is followed by a non-deterministic loop consisting of two options.
In the first option on receipt of a \verb+promise(r,h,v)+ message, the proposer updates \verb+count+ 
(the message counts as a vote), and \verb+hr+ and \verb+hval+ so that they always keep the 
max round identifier and the corresponding value of all received \verb+promise+ messages.
We remark again that \verb+??+ models random channel access.
In the second option we non-deterministically check if a majority has been reached in the current round.
If so, we select the value to send to acceptors for a second votation (done by the learner).
The chosen value is that associated to the maximal round seen so far, i.e., \verb+hval+, if such a value is defined.
If no value has been established yet, i.e. \verb+hval==-1+, the proposer sends its proposed value \verb+myval+ to the acceptors.
In the former case an old value is propagated from one round to another bypassing the value stored in \verb+myval+ (it gets lost). This is needed in order to ensure stability of the algorithm after a choice has been made.
The current round and the selected value are sent to the acceptors using the \verb+baccept+ command.

\paragraph{Acceptor Role}
An acceptor is defined via the following \verb+proctype+:
\begin{lstlisting}
proctype acceptor(int i) {
   short rnd = -1, vrnd = -1, vval = -1;
   short t, t1, j, v, prnd;
   do
     :: rec_a(i,j,v,rnd,vrnd,vval);
     :: rec_p(i,rnd,prnd,vrnd,vval);
     od }
\end{lstlisting}
where the atomic transition \verb+rec_a+is defined as
\begin{lstlisting}
rec_a(i,j,v,rnd,vrnd,vval)=
   atomic { 
          accept??eval(i),j,v ->
          if :: (j>=rnd) -> 
                    rnd=j; 
                    vrnd=j; 
                    vval=v; 
                    learn!i,j,v
             :: else fi; 
          j = 0; v = 0 /* reset */
        }
\end{lstlisting}
and the atomic transition \verb+rec_p+is defined as
\begin{lstlisting}
rec_p(i,rnd,prnd,vrnd,vval)=
  atomic { 
          prepare??eval(i),prnd ->  printf("\nREC\n");
          if :: (prnd>rnd)  -> 
                   promise!prnd,vrnd,vval; 
                   rnd=prnd;
             :: (prnd<=rnd) -> printf("\nSKIP"); fi; 
          prnd = 0 /* reset */
        }
\end{lstlisting}
We use here the following local variables:
\verb+rnd+ contains the current round,
\verb+vrnd+, \verb+vval+ contain resp. the maximal round identifier and the associated value
seen in previous accept messages. Valid values must be greater or equal than zero.
The template consists of a non-deterministic loop with two options.
In the first option on receipt of a \verb+prepare+ message containing a round identifier larger 
than the current one, the acceptor updates \verb+rnd+ and answers to the proposer with a promise message.
The promise contains the values \verb+rnd,vrnd,vval+ used by the proposed to select a value.
In the second option, on receipt of an \verb+accept+  message containing a round identifier larger than the current one, the acceptor updates \verb+rnd,vrnd,vval+ and sends a notification to the learner containing a proposal \verb+vrnd,vval+ for the second votation.
\paragraph{Learner Role} 
The learner role is defined by the following 
proctype
\begin{lstlisting}
proctype learner() {
    short i, j, v;
    byte mcount[MAX];

    do
     :: rec_l(i,j,v,mcount);
    od 
}
\end{lstlisting}
where the transition \verb+rec_l+ is defined as
\begin{lstlisting}
rec_l(i,j,v,mcount)=
   d_step {
        learn??i,j,v ->
        if :: mcount[j] < MAJ -> mcount[j]++
           :: else fi;
        if :: mcount[j] >= MAJ -> printf("\nLEARN\n");
           :: else fi; 
        i = 0; j = 0; v = 0 /* reset */
     }
\end{lstlisting}
A learner keeps track, in the array \verb+mcount+, of the number of received proposals 
values in each round (the counter \verb+mcount[r]+ is associated to round \verb+r+).
We assume that in each round there is a unique accepted value selected by the proposer (no byzantine faults).
This is the reason why we can just count the number of received message in a given round.

%
%
\paragraph{Initial Configuration}
For a fixed number of processes the initial configuration of the system is defined 
using Promela as in the following \verb+init+ command:
\begin{lstlisting}
init 
{ atomic{
   run proposer(1,1); run proposer(2,2);
   run acceptor(0); run acceptor(1); run acceptor(2); 
   run learner(); }; }
\end{lstlisting}
The \verb+atomic+ construct enforce atomic execution of the initial creation of process instances.
In this example we consider three possible proposals that are sent in arbitrary order 
(proposers run in parallel).
Their round identifiers are $1$ and $2$ and the associated values are $1$ and $2$, respectively.
The system has three acceptors with identifiers $0-2$ and a single learner.

\section{Formal Analysis}
We consider here the property of Def. \ref{safetyprop}, which requires that learners always choose the same values. 
Before discussing how to encode the property, we make some preliminary observations.
Since messages are duplicated and seen by any process, we observe that we can restrict 
our model in order to consider a single learner process that is always reacting to 
incoming \verb+learn+ messages. 
The parallel execution of several learners is then modeled via several rounds of a single learner. 
\begin{theorem}\label{abstractlearner}
The safety property of Def. \ref{safetyprop} holds for the model with multiple learners 
if and only if there exist no execution that violates the safety assertion in the model 
with a single instance of the \verb+learner+ proctype.
\end{theorem}
\begin{proof}
Assume that the property is violated in the model containing a single instance of \verb+learner+.
This implies that the instance of the learner performs $k$ iterations in which the learned value 
is always the same and an additional iteration in which the value is distinct from the previous one.
Since the learner only observes incoming messages, we can run the same execution of the protocol with 
distinct instances of the learner process.  
To make the proposition stronger let us assume that each learner learns a single value.

We observe that we just need two distinct instances of the learner process to get a violation of the safety requirement.
We can then run the same instance of the protocol with the single learner.
Since communication is asynchronous,
we can assume that the messages needed in the first $k-1$ are delayed arbitrarily,  and just consider the pairs 
$(r,v)$ and $(r',v')$ learned in steps $k$ and $k+1$, respectively.
The two instances of the learner will learn such pairs and the safety requirement will be violated in the resulting system.

We now assume that safety is violated in the model with multiple learners.
This implies that there exist two distinct pairs $(r,v)$ and $(r',v')$ with $v\neq v'$ 
learned by two distinct learners  (again we assume that each learner learns a single pair).
Again we can run the same execution of the protocol, delay all messages not involving such pairs,
 and let the \verb+learner+ process execute two iterations learning them.
Clearly, the safety assertion will be violated after the votation in the second iteration.
\hfill\qed
\end{proof}
Another important observation is that, since we do not consider byzantine faults,  
it is not possible that two distinct values are proposed in the same round.
Thus, in order to detect a majority we keep an array of counters (one for each round).
The counter for round $r$ is incremented when a message for that round is observed in the 
\verb+learn+ channel. 
Under this assumption we need to show that, once the learner has detected a majority vote for a given value, then the chosen value cannot change anymore.
We define then the \verb+active proctype+ for the following  single learner proctype.
\begin{lstlisting}
active proctype learner() {
 short lastval = -1, id, rnd, lval;
 byte mcount[PROPOSERS];

 do 
 :: read_l(id,rnd,lval,lastval,mcount);
 od
}
\end{lstlisting}
where \verb+read_l+ is defined as follows
\begin{lstlisting}
read_l(id,rnd,lval,lastval,mcount)=
 d_step {
   learn??id,rnd,lval ->
   if 
    :: mcount[rnd-1] < MAJ -> 
         mcount[rnd-1]++;
    :: else 
   fi;
   if 
    ::mcount[rnd-1] >= MAJ   ->
      if :: (lastval >= 0 && lastval != lval) -> 
              assert(false);
         :: (lastval == -1) -> lastval = lval;
         :: else  
      fi
    :: else 
    fi; 
   id = 0; rnd = 0; lval = 0
  }
\end{lstlisting}
The idea is to add an auxiliary variable $\verb+lastvalue+$ in which we store the last learned
value. Every time a new majority is detected the learner compares the corresponding value to
\verb+lastvalue+. An alarm is raised if the two values are not the same.
The alarm is modeled via the assertion $assert(false)$ (or $assert(lastval==v)$).
Since we consider a single learner process that abstracts the behavior of a collection of learners,
we tag the proctype as \verb+active+, i.e., the corresponding process will start together with those specified in the initial configuration (we remove \verb+run learner()+ from \verb+run+).

Apart from the reduction of the number of learners, we can also reduce the number of proposers.
Indeed, to expose violations we just need two proposers proposing distinct values.
This property is formalized in the following statement.
\begin{theorem}\label{abstractproposers}
If for a given value of the parameter $MAJ$ the safety property of Def. \ref{safetyprop} holds for two proposers (with distinct values),
then it holds for any number of proposers.
\end{theorem}
\begin{proof}
Assume a given $k\geq 0$.
By contraposition, we show that if there exists an execution of $k>2$ proposers that violates the assertion in the \verb+single_learner+ code, then there exists an execution with $2$ proposers (distinct rounds and values) 
that violates the assertion.
We consider the first round $r$ that violates the assertion, i.e.,  such that the pair $(r,v')$ obtains a majority observed by the learner for a value $v'$ distinct from the value $v$ stored in \verb+last_val+ (i.e. the value learned in all previous observed votations).
We now have to show that we can construct another execution in which we just need two rounds $r_v$ and $r_{v'}$, 
namely the rounds in which values $v$ and $v'$ have been proposed. 
Let us assume that $r_{v}<r_{v'}$.
The other case is also possible, e.g., for two independent executions involving distinct majorities.
For simplicity we focus on the former case.
To prove that we can define an execution involving rounds $r_{v'}$ and $r_{v}$ that violates the property, 
we need to reason on the history of the protocol phases that produces the necessary majorities.
We start by inspecting a node $n$ that sent the learn message for the pair $(r,v')$.
More specifically, we consider the status of its local variables before the accept message containing $(r,v')$:
\begin{itemize}
\item if they are both undefined, i.e., $n$ did not participate to previous handshakes, 
      $n$ can be reused for an execution involving only $r_{v'}$ and $v'$;
\item if they contain round $r_1$ and value $v$, with $r_1<r$, 
      then we have to show that the history of node $n$ is independent from votations involving round $r$ for the value $v'$.
\begin{itemize}
\item If $n$ has never voted for value $v'$ (i.e. the value $v'$ has never been stored in its local state), 
      then we can inspect the history until the first vote done by $n$ for the value $v$.
      If the associated round is $r_v$, we can simply build an execution in which node $n$ sends the learned 
      message $(r_v,v)$. If the round number is different from $r_v$, then the previous local state has undefined 
      values for the variables (since we assume that this is the first votation).
      Thus, we can build an execution in which node $n$ receives the proposal $(r_v,v)$ directly from the proposer $r_v$.
\item If $n$ has voted for value $v'$ in a round $r_1<r$ but the value has not reached a majority, i.e., the learner 
      never observed a majority for $(r_1,v')$, then the vote of node $n$ plays no role for the election of $v'$.
      Thus, we can apply the same reasoning as in the previous point, in order to move back to a state from which we can 
      extract an execution involving only $r_v$ and $v$.  
\end{itemize}   
\end{itemize}
\qed
\end{proof}
We conjecture that a similar finite-model result holds, for the considered model, even for defining a bound on the number of acceptor.
Namely, we believe that studying the protocol for a small number of acceptors defining a potential partitioning is sufficient 
for proving the protocol correct. We leave the proof of this claim for an extended version of the paper.
\paragraph{Reducing the Search Space}
Apart from finite-model properties, it is also possible to apply a number of heuristics to reduce the state space of the automata
associated to the protocol roles.
The first heuristics consists in resetting all locally used variables at the end of atomic steps.
This, we increase the probability that the automaton rule associated to the atomic step returns to an existing state
(unless global variables have been updated by the rule).
 
The second heuristics is strictly related to the channel representation.
Since we always use the random read \verb+??+ for message reception, 
we can use the ordered insertion \verb+!!+ to send messages instead of the 
FIFO version \verb+!+.
Conceptually, there is no difference (read operations remain unordered),
however we reduce the number of state representation by keeping only 
representative elements in which channel contents are always ordered lists.

\begin{table}[t]
{\scriptsize
\begin{center}
\begin{tabular}{|r|r|r|r|r|r|r|r|r|}
\hline
Prop & Acc & Max & Maj & Time (sec)  & App  & States           & Unsafe & OutOfMem\\
\hline
 2  & 2  & 3   & 1   &   0.006    &      &          346      &  X & \\  
 2  & 2  & 3   &  2  &   0.007  &      &            824   &    &      \\  
 2  & 2  & 4   &  2  &   0.006  &      &            790    &    &     \\
 2  & 3  & 3   &  1  &   0.014  &      &            2195  &  X & \\
 2  & 3  & 6   &  2  &   0.285  &      &           65091   &    &   \\
 2  & 4  & 3   &  1  &   0.087  &      &           20045  &  X & \\
 2  & 4  & 3   &  2  &   0.3    &      &           52281  &  X & \\
 2  & 4  & 3   &  3  &   4.06   &      &          979118   &    &    \\
 2  & 4  & 6   &  3  &   5.81   &      &          794775  &    &     \\
 2  & 4  & 7   &  3  &   5.81   &      &          754072  &    &    \\
 2  & 4  & 8   &  3  &   5.97   &      &          744224  &    &    \\
 2  & 4  & 9   &  3  &   6.00   &      &          744224   &    &    \\
 2  & 5  & 3   &  1  &   0.995  &      &          151550   &  X &   \\
 2  & 5  & 3   &  2  &   2.55   &      &          377295   &  X &   \\
 2  & 5  & 3   &  3  &  57.20   &      &         7807712   &    & X \\
 2  & 5  & 10  &  3 &   61.5    &      &         5186311   &    & X \\
 2  & 5  & 12  &   3 &   57.2   &      &          4708912  &    & X \\
 2  & 5  & 15  &   3 &   56.1   &      &           4152603 &    & X \\
 2  & 5  & 20  &   3  &  56     &      &           3476531 &    & X \\
 2  & 5  & 10  &   3  &  317    & X   &    53*$10^6$        &    &    \\
 2  & 6  & 12  &   2   &  17    & X    &   3*$10^6$         &    X& \\
 2  & 6  & 12  &   3   &  35    & X    &   7*$10^6$         &    X& \\
 \hline
 \end{tabular}
\end{center}
}
 \caption{Experimental results with finite protocol instances.}
 \label{experiments}
 \end{table}
\section{Experimental Analysis}
In our experiments we consider the number of proposer and acceptors as distinct parameters, 
the other parameters being the maximum size of channels (communication is asynchronous),
and the maximum number of faulty processes allowed in the system.
We use Spin as back-end solver to explore the values of the parameters and 
to extrapolate the minimal constraints for ensuring the correctness of the protocol.
In a first series of experiments (with ordered send operation $!!$ instead of $!$) 
we considered the parameters in Table \ref{experiments}.
With two proposers and two acceptors we detect a violation with majority 1 and
verify the safety property with full state exploration with majority 2.
Similarly, we detect a violation with three acceptors and majority 1
and no violations with majority 2. Here we consider increasing size channels.
The state space stabilizes, as expected, for buffer of size 6.
With 4 acceptors we prove correctness for majority 3 and the state space stabilizes, 
as expected, with channel of size 8. 
We remark that we obtain almost a half reduction with respect to a model with FIFO send operation $!$
(e.g. for $3$ acceptors and channel size $6$, the reachable states are 172868 with $!$ and 65091 with $!!$).
With 5 acceptors the state space becomes unmanageable with full state exploration.
We can still apply approximated search (bitstate hashing) to get violations for quorums with less than 3 processes and have an estimation of the correctness for larger quorums.
The experiments confirm the hypothesis for the correctness of the protocol.
From the minimal constant $MAJ$ that show no violation we can extrapolate the 
number of admitted faulty processes.
Interestingly,  Spin is still very effective in finding counterexamples as shown by 
the experiments in Fig. \ref{experiments}.
 
\section{Optimizations via Model Transformations}
In this section we present two optimizations, given in form of model transformations, that help in reducing 
the search space of our models.

The first optimization can be applied in the intial phase of the protocol of a proposer.
Specifically, we can send prepare messages atomically to all processes, i.e., perform an atomic
broadcast transition instead of interleaving send of prepare messages with actions of other processes.

The second optimization is related to the method used to detect a majority by proposers.
Counting up messages using counters on reception of promise messages has a major drawback, i.e., 
the introduction of a number of intermediate states proportional to the steps needed to reach a majority 
(every time we receive a message we increment a counter).
To eliminate these auxiliary states, we need a new type of guards that are able to count the occurrences of a 
given message in a channel. We will refer to them as counting guards.
When used in a conditional statement they would allow us to atomically check if 
a quorum has been reached for a specific candidate message, i.e., to implement atomic  a sort of 
quorum transitions.

If we apply the two transformations to our Promela specification of proposers, 
we obtain a much simpler process skeleton in which an atomic broadcast is followed by a \verb+do+
loop consisting of a single  rule that models a quorum transition.
To implement these steps we use derived Promela rules built via the \verb+atomic+ and \verb+d_step+ construct.
The \verb+d_step+ construct is particularly useful in that it transforms a block of transitions into a single not interruptible step, 
i.e., it can be used to add new instructions to Promela.
Only the first instruction of a \verb+d_step+ block can be a blocking operation, e.g., a read from a channel.
We discuss the new Promela model in the rest of the section.
\paragraph{Atomic Broadcast and Quorum Transitions}
In the new specification we introduce a new type for messages \verb+mex+ 
used to inspect the content of the channel \verb+prepare+.
\begin{lstlisting}
typedef mex{
	byte rnd;
	short prnd;
	short pval;
}
\end{lstlisting}
The new proposer definition is structured as follows.
\begin{lstlisting}
proctype proposer(short crnd; short myval) {
 short aux, hr = -1, hv = -1;
 short rnd;
 short prnd,pval;
 byte count=0,i=0;
 mex pr;

 d_step{ bprepare(crnd); }
 do
 :: qt(i,pr,count,hr,hv,myval,crnd,aux);
 od
}
\end{lstlisting}
As mentioned before, the first step consists in atomically broadcasting prepare messages. 
The process then enters a loop in which it only executes the derived \verb+qt+ rule.
The \verb+qt+ rule is defined via the \verb+atomic+ construct and two additional subrules as follows.
\begin{lstlisting}
qt(i,pr,count,hr,hv,myval,crnd,aux) =  
  atomic{
   occ(i,pr,count,hr,hv,crnd);
   test(count,hr,hv,myval,crnd,aux);
   hv= -1;  hr = -1; count =0; aux=0; 
  }
\end{lstlisting}
\verb+occ+ is a rule that counts the number of occurrences of messages with round identifier equal to \verb+rnd+.
\verb+test+ is used to detect a majority and broadcast \verb+accept+ messages.

The \verb+occ+ procedure on a generic channel $ch$ and message $m$ is based on the following idea. 
We consider the $ch$ channel as a circular queue.
We then perform a for-loop as many times as the current length of the channel.
At each iteration we read message $m$ by using the FIFO read operation $?$ and reinsert it 
at the end of the channel so as to inspect its content without destroying it.
We then increment an occurrence counter $count$ if the message $m$ contains the 
proposed round identifier.
The other fields of the messages are inspected as well in order to search for the 
value of the promise containing the maximal round identifier.
At the end of the for-loop we can test the counter to fire the second part of the transition
in which we test if the majority has been reached.
The promela code for the \verb+occ+ subrule is defined then as follows.
\begin{lstlisting}
occ(i,pr,count,hr,hv,crnd) =
  d_step{
   do 
   i=0; 
   :: i < len(promise) ->
     promise?pr; promise!pr; 
     if 
      ::pr.rnd==crnd->
       count++;
       if
        ::pr.prnd>hr->
           hr=pr.prnd; hv=pr.pval;
       ::else
      fi;
      ::else
     fi;
     i++;
   :: else -> 
     pr.prnd=0; pr.pval=0; pr.rnd=0; i=0;
     break;
   od;
   } 
\end{lstlisting}
The instructions in the outermost \verb+else+ branch are just resets of local variables.

An accept message is broadcasted to the acceptors when majority is detected.
This is implemented in the \verb+test+procedure defined as follows.
\begin{lstlisting}
test(count,hr,hv,myval,crnd,aux) =
  if 
    ::count>=MAJ -> 
      aux=(hr<0 ->myval : hv); /* conditional expression */
      baccept(crnd,aux); 
      break;
    ::else 
   fi;
\end{lstlisting}
We remark that the quorum transition rule is executed in a single deterministic step, i.e.,
it does not introduce any intermediate state.
If the proposer detects a majority then the effect is to update the \verb+accept+ channel.

We note that we still consider possible interleaving between different quorums since 
acceptors are free to reply to proposers in any order.
More specifically, the reason why the transformation preserves all interleavings of the 
original model is  explained by the following proposition.
\begin{proposition}
The model obtained by applying the transformation based on atomic broadcast and quorum transitions 
is equivalent to the original model with respect to set of messages exchanged by proposers and acceptors.
\end{proposition}
\begin{proof}
In the first model a proposer broadcasts requests non-atomically and then moves to a control state in which
it waits for replies and for detecting a majority on some round.
At each reply a proposer updates the local variables until majority is reached and then updates the \verb+accept+ channels.
We now show that the modified model has the same effect on the channels in terms of read and write operations.
We prove the claim in two steps.
We first show how to consider only atomic broadcast and then how to replace receptions with quorum transitions.
\begin{itemize}
\item
Let us consider a computation $\gamma_1 R_i S_j\gamma_2$ of the first model in which $R_i$ is a reception performed
by proposer $P_i$ and $S_j$ is a send issued by proposer $j$. 
Since the communication model is asynchronous, reception $R_i$ can be permuted with $S_j$.
Indeed, $S_j$ cannot disable $R_i$ and $S_j$ does not depend on $R_i$.
Thus,  $\gamma_1 R_i S_j\gamma_2$ has the same effect on channels.
By iterating this permutation rule, we can group all sends issued by a given process 
in a contiguous subsequence of actions and then replace them with an atomic broadcast step.
Restricting the interleavings to the above mentioned types of interleavings does 
not change the effect on the channels.
\item
We now consider the second type of transformation.
We first note that, in general, receive operations cannot be grouped together.
Indeed, in the first model each receive operation could modify the current state of a proposer 
by updating $count$.
We observe however that the effect of a single receive operation followed by an update and a test on $count$
is the same as the effect of a quorum transition applied directly on the channel.
In other words a computation of the form 
$\gamma_1 R_1 \gamma_2 R_2 \ldots \gamma_k R_k \gamma_{k+1}$
where $R_i$ is a receive operated by proposer $P_i$ as specified in the first model, 
is mimicked by an equivalent computation of the form  
$\gamma_1 Qt \gamma_2 Qt \ldots \gamma_k Qt \gamma_{k+1}$
in which a reception with local state and \verb+accept+ channel updates 
is replaced by an application of a quorum transition $Qt$ whose update is 
limited to the \verb+accept+ channel.
This property is ensure by the fact that, in case of failure, a quorum transition goes back to 
the beginning of the \verb+do+ loop and can be fired again in any other step. 
\end{itemize}
\end{proof}
\paragraph{Prepare Channels as Read-only Registers}
Another fundamental optimization can be applied in the acceptor specification.
The key point here is to consider the propose channel as a sort of message store instead of a standard channel,
i.e., a channel in which message are persisent.
To obtain this effect, it is enough to turn the read operation \verb+prepare??eval(id),rnd+ that removes the message from
the channel into the read command \verb+prepare??<eval(id),rnd>+ that does not remove the message from the channel.
The new acceptor specification is defined next.
%
\begin{lstlisting}
proctype acceptor(int id) { 
 short crnd = -1, prnd = -1, pval = -1;
 short aval,rnd;
 do
  ::d_step { 
    prepare??<eval(id),rnd> ->
    if 
     :: (rnd>crnd)  -> crnd=rnd;
     :: else 
   fi; 
   rnd = 0
  }
  :: d_step { 
    accept??eval(id),rnd,aval ->
      if 
       :: (rnd>=crnd) ->
        crnd=rnd; 
        prnd=rnd;
        pval=aval;
        learn!!id,crnd,aval; 
       :: else 
     fi; 
     rnd = 0; aval = 0;
   }
 od
}
\end{lstlisting}
It is important to notice that, by definition, acceptors never process more than once messages with a certain round identifier.
Therefore, leaving the messages in the channel does not modify the traces of acceptors.
This optimization can be pushed forward by letting proposer to put a single message for each round in the \verb+prepare+ channel
instead of one message per process. In other words, prepare messages can be viewed as read-only registers accessible to acceptors.

\section{Experimental Analysis of the Optimized Model}
In this section we discuss the experimental results obtained on the optimized model with the parameters 
shown in Table \ref{experiments1}.
\begin{table}[t]
{\scriptsize
\begin{center}
\begin{tabular}{|r|r|r|r|r|r|r|r|r|}
\hline
Prop & Acc & Max & Maj & Time (sec)  & App  & States           & Unsafe & OutOfMem\\
\hline
 2  & 2  & 4   &  2  &   $<$0.01  &      &            61   &    &      \\  
 2  & 3  & 6   &  2  &   $<$0.01  &      &           926   &    &   \\
 2  & 4  & 8   &  3  &   0.02   &      &          4421   &    &    \\
 2  & 5  & 10   &  3  &  0.4   &      &         91956   &    &  \\
 2  & 6  & 12  &  4  &  2.74  &           &          473261   &     &  \\
 2  & 7  & 14  &  4  &  71.9  &           &          9762358 &   &  \\
 2  & 8  & 16  &  5  &  799  &           &           28208534 &  & X \\
 2  & 8  & 16  &  5  &   52.3 &    X       &           7525860  &  &\\
 2  & 8  & 16  &  4  &   60.3     &  X         &      8550176   &  &\\
 3  & 2  & 6   &  2  &  0.01  &           &          987 &   &\\
 3  & 3  & 9   &  2  &  0.35  &           &          104761 &   &\\
 3  & 4  & 12  &  3  &  7.3  &           &          1339802 &   &\\
 3  & 5  & 15  &  3  &  940  &           &          30813015 &   & X\\
 3  & 5  & 15  &  3  &   49    &     X     &          8090684  &   &\\
 2  & 5  & 15  &  2  &   48.2  &    X      &      7383348     & X &\\
 4  & 2  & 8   &  2  &   0.07  &           &          16457  &  &\\
 4  & 3  & 12  &  2  &  48.1  &           &          9547976 &  & \\
 4  & 4  & 16  &  3  &   657  &           &       29450661 &   & X\\  
 4  & 4  & 16  &  3  &    51.8 &    X     &        8128276 &   & \\
 4  & 4  & 16  &  2  &     49.1    &    X     &     7870336   & X &\\  
 5  & 2  & 10  &  2  &    1.43    &            &        270237   &   & \\
 5  & 3  & 15  &  2  &    $>$3600     &               &     $>$1.22*$10^8$         &    & X\\
 5  & 3  & 15  &  2  &   50          &    X          &     7949467     &       & \\
 5  & 3  & 15  &  1  &    2.17       &   X          &      340445     &   X     &\\
 6  & 2  & 12  &  2  &     25.9     &               &    4397176    &       &\\
 7  & 2  & 14  &  2  &    558          &               &    71298752   &       &\\
 8  & 2  & 16  &  2  &    2370        &               &    1.4984*$10^8$ &     & X\\
 8  & 2   & 16  & 2  &     70.4        &   X            &    8045888 &       &\\ 
 8  & 2   & 16  & 1  &    17.4         &  X               &  2189799 & X  &\\         
 \hline
 \end{tabular}
\end{center}
}
 \caption{Experimental results with finite protocol instances.}
 \label{experiments1}
 \end{table}
As their number of processes increases we observe a fast growing number of states due to the asynchronous nature of the model.
When our PC revealed to be not sufficiently powerful to exhaustively look for the correctness of the system in every single possible state and an out-of-memory error was generated, an approximate search option was used (App stands for -DBITSTATE option given for compilation process along with 3 bits per state and default hash-table size chosen for the verification phase). In that case Spin bitstate hashing turned out to be useful when quorums where not enough to reach a majority and a counterexample was given almost in any case. The correct approximate case instead has little meaning because, even increasing the internal hash table size and using more bits to represent a state, the hash factor remains well below a good states coverage target. In any case, the abstractions used to reduce the state-space proved to be successful to analyse a few processes and  to tackle the complexity of distributed system verification. When approximate search was not needed we compressed the state descriptors using the appropriate Spin option to fully verify more effectively. Spin was run on a 16GB machine using a presetting of 4GB for the state vector. All experiments have been done with a PC equipped with an Intel i7-4700HQ quad-core with hyper-threading enabled by default.
We remark that we obtain almost a half reduction with respect to a model with FIFO send operation $!$
With 8 acceptors the state space becomes unmanageable with full state exploration.
We can still apply approximated search (bitstate hashing) to get violations for quorums with less than 3 processes and have an estimation of the correctness for previously unmanageable configurations.
The experiments confirm the hypothesis about the correctness of the protocol.
From the minimal constant $MAJ$ that show no violation we can extrapolate the 
number of admitted faulty processes. We also increased step by step the $MAX$ parameter to get to its stabilization to confirm our hypothesis about its size. 
Interestingly,  Spin is still very effective in finding counterexamples as shown by 
the experiments in Fig. \ref{experiments1}.
Our models are available here: \url{http://www.disi.unige.it/person/DelzannoG/PAXOS/GANDALF_14/}.

%

\section{Related Work}
\label{related}
Formal specification in temporal and first order logic, TLA, of Paxos and its variants together with automatically checked proofs are given in \cite{L11}.
Differently from the TLA-based approach, our analysis is based on a software model checking approach 
based on automata combined with abstractions and heuristics (e.g. counting guards).
The efficiency of model checking message passing protocols is studied in \cite{bokor2010efficient}, which compares different semantics for asynchronous communication (e.g. one in which special delivery events are used to move messages from output to input buffers and another in which messages are directly placed in input buffers).
The authors conclude the article with some results obtained by applying the discussed message passing models to Paxos, with two proposers and up to four acceptors. Depending on the specific model and property being considered, the state space varies from about $5\times 10^{4}$ states up to $1.7 \times 10^{6}$. An approach for bounded model checking asynchronous round-based consensus protocols is proposed in \cite{tsuchiya2008using}, where the authors exploit invariants and over-approximations to check agreement and termination properties on two variants of Paxos. In \cite{JKSVW13,GKSVW14}, the authors consider (parameterized models of) fault tolerant distributed algorithms.
They propose an approach specific to this class of protocols and consider more types of faulty processes than we do, like Byzantine failures.
Differently from  \cite{JKSVW13,GKSVW14}, we focus our attention on optimisations and code-to-code transformations
that can help in reducing state space for finite-state instances of protocols in the same class.
Reduction theorems and abstractions for other examples of distributed systems (e.g. mutual exclusione protocols) 
are considered in \cite{EN03,NT13}.  Automata-based models of broadcast communication  has recently been studied in \cite{DSZ10,DSZ11,DSTZ12,DT13b,DST13}. 

Concerning other model checking approaches, in \cite{graphite14}, we focused our attention on the comparison of the model checkers Spin and Groove, based on graph transformation systems, taking Paxos as a common case-study. 
In this paper we focus our attention on the design choices  and protocol properties that we applied to obtain the Promela specification. The Groove model studied in \cite{graphite14} can be viewed as a declarative specification of Paxos that require a preliminary abstraction  step in the modeling phase.
We apply here code to code transformations and optimizations to reduce the state space generated by exhaustive analysis via Spin. This method, that requires human ingenuity, yields results that are comparable, in terms of matched states and execution time, to those obtained with Groove.

\section{Conclusions}
\label{conclusions}
In this paper we presented a formal model for the Paxos algorithm given in terms of finite state automata described in the high level language Promela. Reasoning on the size of the automata and reachability states produced by Spin, we managed to define different types of optimizations, e.g., based on a new type of guards that atomically inspect the content of a channel (counting guards). We also consider two reduction properties that can be used to limit some of the unbounded dimension of the specification (number of proposers and learners). 
Our experiments show that the combination of abstractions, heuristics, and model checking can be used to analyze challenging examples of distributed algorithms even for large number of process instances.
As future work, it would be interesting to generalize the transformation patterns (based on atomic broadcast and quorum transitions) to a more general
class of systems and to provide a complete cut-off result to reduce parameterized verification to finite-state verification.

\bibliographystyle{eptcs}
\bibliography{biblio}
\end{document}